\documentclass[]{interact}

\usepackage{epstopdf}
\usepackage[caption=false]{subfig}

\usepackage[numbers,sort&compress]{natbib}
\usepackage{xcolor}

\bibpunct[, ]{[}{]}{,}{n}{,}{,}
\makeatletter
\def\NAT@def@citea{\def\@citea{\NAT@separator}}
\makeatother

\theoremstyle{plain}

\theoremstyle{definition}

\theoremstyle{remark}

\usepackage{url} 

\begin{document}


\title{Exploring utilization of generative AI for research and education in data-driven materials science}

\author{
\name{Takahiro~Misawa\textsuperscript{a}, Ai Koizumi\textsuperscript{b}, Ryo Tamura\textsuperscript{b,c}, and Kazuyoshi Yoshimi\textsuperscript{a}}
\affil{\textsuperscript{a}Institute for Solid State Physics, University of Tokyo, 5-1-5 Kashiwanoha, Kashiwa, Chiba 277-8581, Japan; \textsuperscript{b}National Institute for Materials Science, Namiki, Tsukuba-shi, Ibaraki, 305-0044, Japan;
\textsuperscript{c}Graduate School of Frontier Sciences, University of Tokyo, 5-1-5 Kashiwanoha, Kashiwa, Chiba 277-8581, Japan}
}

\maketitle

\begin{abstract}
Generative AI has recently had a profound impact on various fields, including daily life, research, and education.  
To explore its efficient utilization in data-driven materials science, 
we organized a hackathon---AIMHack2024---in July 2024. In this hackathon, researchers from fields such as materials science, information science, bioinformatics, and condensed matter physics worked together to explore how generative AI can facilitate research and education. 
Based on the results of the hackathon, this paper presents topics related to (1) conducting AI-assisted software trials, (2) building AI tutors for software, and (3) developing GUI applications for software. 
While generative AI continues to evolve rapidly, this paper provides an early record of its application in data-driven materials science and highlights strategies for integrating AI into research and education.
\end{abstract}


\section{Introduction}
The emergence of generative AI, such as ChatGPT~\cite{ChatGPT}, has had a significant impact on various fields, including materials science, where it has been used for tasks such as data analysis, automated literature review, and computational modeling. In July 2024, to explore efficient ways to integrate generative AI into research and education in materials science, specifically in automated data analysis, literature review, and software development, we organized a hackathon, AIMHack2024~\cite{AIMHack2024}.
Approximately 30 participants, ranging from students to industry researchers, took part in the hackathon. The participants came from diverse backgrounds, including materials science, computational science, bioinformatics, and condensed matter physics, which fostered interdisciplinary collaboration.

The workshop was designed to accommodate both beginners and advanced users. For beginners, the guided demonstrations introduced fundamental steps for using selected software from a curated list on a portal site such as MatDaCs (MaterialsDataCommons)~\cite{MatDaCs}.
Through MatDaCs, participants could access and use various computational materials science applications, including machine learning frameworks. Participants were encouraged to perform basic calculations with the selected software, gaining practical experience under the supervision of generative AI.
Advanced users, on the other hand, were given the flexibility to select and independently explore software of their choice, following a structured workflow similar to that used by beginners. This approach allowed them to investigate more advanced features and customize their usage according to their research interests.

The primary emphasis of the hackathon was to enhance participants' ability to critically evaluate AI-generated content. Participants were encouraged to systematically verify AI-generated responses by cross-referencing them with established literature and their domain knowledge. They also tested the AI-generated code and tutorials by running the recommended software, assessing both functionality and accuracy. This exercise was designed not only to enhance technical proficiency but also to foster critical thinking about the reliability and applicability of AI-generated content.

The collaborative efforts in the hackathon aimed to refine participants' strategies for using AI tools in their research activities, facilitating AI integration into research workflows and practical scientific applications.
The key results of the hackathon included 
the implementation of AI-assisted software trials using ImageJ~\cite{ImageJ} for automated image analysis, the development of an AI tutor for PHYSBO~\cite{MOTOYAMA2022108405}, a Bayesian optimization tool, using MyGPTs, and the creation of a graphical user interface (GUI) to enhance the usability of PHYSBO. These results highlight the increasing impact of generative AI in advancing materials science research and education.

We note that these three topics are closely related to the typical workflow in materials science 
research. First, image recognition of experimental results is a crucial first step for data accumulation, 
and mastering the use of ImageJ directly contributes to efficient data acquisition.
Second, Bayesian optimization is a widely used method for determining the next experimental conditions based 
on accumulated data, and gaining a deeper understanding of PHYSBO through an AI tutor is important for 
improving research quality. Third, a GUI is useful for executing Bayesian optimization in a user-friendly 
manner, and our demonstration of a simple method for GUI development provides practical insights for a broad 
range of developers. Thus, the three topics addressed in this paper correspond to essential components of the research process in materials science.

This paper is organized as follows. Section \ref{sec:review} presents AI-assisted software trials using ImageJ. Section \ref{sec:AIassist} discusses the development of an AI tutor using MyGPTs for PHYSBO. Section \ref{sec:GUI} describes the creation of a GUI for PHYSBO with AI support. Finally, Section \ref{sec:Summary} provides a summary of the findings and discusses future directions for AI in materials science.

\section{AI-assisted trial use of software}
\label{sec:review}
In this section, we present an example of how ChatGPT can be used to assist in the trial use of software for materials science. During the hackathon, we chose the image analysis software ImageJ~\cite{ImageJ} from the list of software available on MatDaCs~\cite{MatDaCs} as an example. Since ImageJ can count objects in an image, we decided to explore this feature with it.

\begin{figure}[tb]
   \centering
   \includegraphics[width=12.5cm]{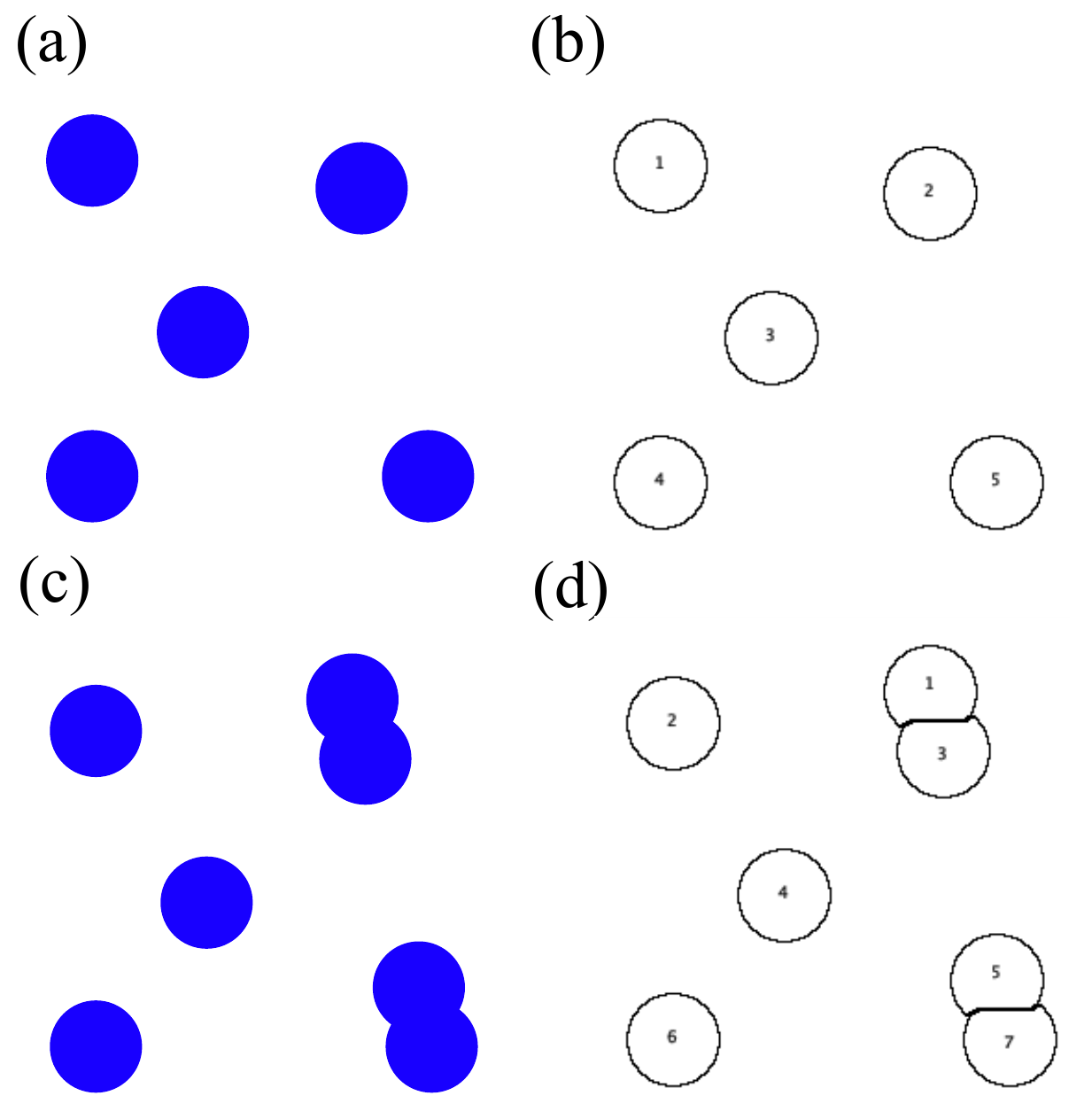}
   \caption{Sample images: (a) Circles without overlap and (c) Circles with overlap. ImageJ analysis results: (b) Counting result for non-overlapping circles and (d) Counting result for overlapping circles.
   }
   \label{fig:ImageJ}
\end{figure}

As an example, we prepared an image with several circles, as shown in Fig. 1(a). We then uploaded this image to ChatGPT and entered the following prompt:
\begin{quote}
    \textit{
"How can I count the number of circles in this image using ImageJ?".}
\end{quote}
ChatGPT generated the following step-by-step instructions~\footnote{We used ChatGPT 4o and ImageJ 1.54g in this tutorial.}:

\begin{enumerate}
    \item \textbf{Open the Image}
    \begin{itemize}
        \item Launch ImageJ.
        \item Select \texttt{File → Open...} from the menu and open the target image.
    \end{itemize}

    \item \textbf{Convert to Grayscale}
    \begin{itemize}
        \item Select \texttt{Image → Type → 8-bit} from the menu to convert the image to grayscale.
    \end{itemize}

    \item \textbf{Adjust the Threshold}
    \begin{itemize}
        \item Select \texttt{Image → Adjust → Threshold...}.
        \item Use the sliders to select only the circles (they should appear in red).
        \item Click "Apply."
    \end{itemize}

    \item \textbf{Convert to Binary Image}
    \begin{itemize}
        \item Select \texttt{Process → Binary → Make Binary} to create a black-and-white image.
    \end{itemize}

    \item \textbf{Perform Particle Analysis}
    \begin{itemize}
        \item Select \texttt{Analyze → Analyze Particles...}.
        \item Set an appropriate "Size" range (e.g., \texttt{100-Infinity} to filter out noise).
        \item Check \texttt{"Display results"}, \texttt{"Show outlines"}, and \texttt{"Summarize"}.
        \item Click "OK" to perform the analysis.
    \end{itemize}

    \item \textbf{View the Results}
    \begin{itemize}
        \item The results window will display the count of detected circles.
    \end{itemize}

\end{enumerate}

By following these instructions and operating ImageJ, we were able to count the circles, as shown in Fig. 1(b).
Typically, when using such software, one begins by reading manuals and learning how to use it, which can take a considerable amount of time. However, using generative AI in this way, we can significantly reduce the effort required for learning the software, which presents a major advantage. In fact, when we demonstrated this result, many participants were greatly amazed by how convenient generative AI is.

Furthermore, as a more advanced test, we examined whether ImageJ could distinguish overlapping objects. For this test, we prepared an image with several overlapping circles, as shown in Fig. 1(c). We then uploaded this image to ChatGPT and entered the following prompt:
\begin{quote}
    \textit{
``How can I count individual particles when they overlap, as shown in the attached image?".}
\end{quote}
ChatGPT generated the following step-by-step instructions:
\begin{enumerate}
    \item \textbf{Open the Image}
    \begin{itemize}
        \item Launch ImageJ.
        \item Select \texttt{File → Open...} from the menu and open the target image.
    \end{itemize}

    \item \textbf{Convert to Grayscale}
    \begin{itemize}
        \item Select \texttt{Image → Type → 8-bit} from the menu to convert the image to grayscale.
    \end{itemize}

    \item \textbf{Adjust the Threshold}
    \begin{itemize}
        \item Select \texttt{Image → Adjust → Threshold...}.
        \item Use the sliders to highlight only the circles (they should appear in red).
        \item Click "Apply."
    \end{itemize}

    \item \textbf{Convert to Binary Image}
    \begin{itemize}
        \item Select \texttt{Process → Binary → Make Binary} to create a black-and-white image.
    \end{itemize}

    \item \textbf{Apply Watershed Segmentation}
    \begin{itemize}
        \item Select \texttt{Process → Binary → Watershed}.
        \item This will separate touching circles by adding dividing lines.
    \end{itemize}

    \item \textbf{Perform Particle Analysis}
    \begin{itemize}
        \item Select \texttt{Analyze → Analyze Particles...}.
        \item Set an appropriate "Size" range (e.g., \texttt{100-Infinity} to filter out noise).
        \item Check \texttt{"Display results"}, \texttt{"Show outlines"}, and \texttt{"Summarize"}.
        \item Click "OK" to perform the analysis.
    \end{itemize}

    \item \textbf{View the Results}
    \begin{itemize}
        \item The results window will display the number of detected circles.
        \item A new image will appear with outlined particles.
    \end{itemize}

\end{enumerate}

From these results, we learned that the {\it watershed} function can be used to detect overlapping areas.
By following these instructions and operating ImageJ, we confirmed that the watershed function successfully separated overlapping regions, allowing for accurate object counting, as shown in Fig. 2(d).
In a standard approach before the emergence of generative AI, finding such a feature required reading the manual and searching for the relevant section. However, we found that this effort could be significantly reduced by asking generative AI.

This experiment demonstrated that generative AI can be an effective tool for learning how to use unfamiliar software.
Traditionally, learning how to use software typically required conducting web searches and gathering information from relevant websites. However, generative AI is emerging as an alternative to this process.
Based on this tutorial, several participants successfully used software they had not used before.

\section{Building an AI Tutor for PHYSBO with MyGPT}
\label{sec:AIassist}
Recent advances in generative AI have accelerated automated code maintenance in software development ~\cite{bubeck2023sparks, chen2021evaluating}. In particular, the emergence of Large Language Model (LLM)-based tools for code completion and debugging assistance has contributed to developer productivity ~\cite{zhang2023practices}: tools such as GitHub Copilot~\cite{githubcopilot} and Amazon CodeWhisperer~\cite{amazoncodewhisperer} can understand the intent of a program, suggest appropriate code, and complete it accordingly~\cite{Imai, pearce2021asleep}.
This technological advance is moving beyond code completion to a deeper understanding of the semantics of code. In the future, AI may be able to integrate software specifications, manuals, and source code to automatically generate tutorials and documentation for users.

In this section, we explore the above possibility of using MyGPT to automatically generate tutorials for a specific piece of software.
MyGPT is a tool that allows users to build customized conversational AI based on pre-provided information. 
The goal of this section is to evaluate the usefulness of MyGPT in generating tutorials when relevant information is provided in advance. 
Specifically, we assess MyGPT's ability to generate tutorials in terms of accuracy.
Based on the results, the potential for future AI-based document generation is discussed.

As our target software, we selected PHYSBO~\cite{MOTOYAMA2022108405}, a Bayesian optimization package. Bayesian optimization is a method that uses Gaussian processes as a regression model to estimate the objective function from explanatory variables and suggest those predicted to yield the best objective function value. 
One of the key features of PHYSBO is its ability to propose multiple candidates and appropriately select an acquisition function, including Thompson sampling (TS), expected improvement (EI), and probability of improvement (PI). Furthermore, the computational cost, which typically requires $O(N^3)$ for the number of training data $N$, has been reduced to $O(N)$.

To evaluate the effectiveness of MyGPT in generating accurate tutorials and explanations for PHYSBO, we provided it with various sources of reference information. Initially, we configured MyGPT with a system prompt specifying its role as an AI assistant specializing in PHYSBO. The following official sources were used as reference materials:
\begin{itemize}
    \item \textbf{GitHub repository}: {https://github.com/issp-center-dev/PHYSBO}
    \item \textbf{Official paper}: {https://www.sciencedirect.com/science/article/pii/S0010465522001242}
    \item \textbf{User manual}: {https://issp-center-dev.github.io/PHYSBO/manual/master/en/index.html}
\end{itemize}

Using this setup, we tested the accuracy of MyGPT by asking it the following question:
\begin{quote}
    \textit{``What acquisition functions does PHYSBO support?''}.
\end{quote}
However, MyGPT initially provided an incorrect response, claiming that PHYSBO implements the Upper Confidence Bound (UCB) acquisition function, which is not implemented.
Additionally, MyGPT failed to mention PHYSBO's support for multi-objective optimization.

\begin{figure}
    \centering
    \includegraphics[width=1 \linewidth]{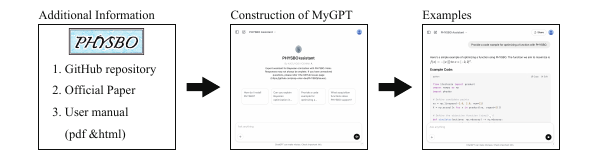}
    \caption{Overview of the process for generating tutorials for PHYSBO using MyGPT. MyGPT was initially tested only with URLs to PHYSBO’s official documentation, followed by an improved setup where full documentation PDFs were provided. }
    \label{fig:physboAI}
\end{figure}

To improve the accuracy of MyGPT's responses, we uploaded the official PHYSBO documentation and the research paper as PDF files. 
We then repeated the same query, resulting in a significantly improved and accurate response. MyGPT correctly listed PHYSBO’s available acquisition functions:

\textbf{For Single-Objective Optimization}
\begin{itemize}
    \item \textbf{Thompson Sampling (TS)} – Selects the input that maximizes a sampled regression function from the posterior distribution.
    \item \textbf{Expected Improvement (EI)} – Selects the input where the expected value of improvement over the current best value is maximized.
    \item \textbf{Probability of Improvement (PI)} – Selects the input with the highest probability of exceeding the current maximum objective function value.
\end{itemize}

\textbf{For Multi-Objective Optimization}
\begin{itemize}
    \item \textbf{Thompson Sampling (TS)} – A multi-objective extension of Thompson Sampling, which samples from multiple Gaussian processes.
    \item \textbf{Hypervolume-Based Probability of Improvement (HVPI)} – Measures the probability of improving the non-dominated region in multi-dimensional objective space.
    \item \textbf{Expected Hypervolume Improvement (EHVI)} – Estimates the expected improvement in hypervolume for Pareto-front optimization.
\end{itemize}

Furthermore, MyGPT accurately noted that these acquisition functions balance exploration and exploitation in Bayesian optimization.
It also correctly stated that PHYSBO allows users to specify the acquisition function via the \texttt{score} argument in its optimization functions.

The results indicate that providing URLs alone was insufficient to ensure accurate responses, since MyGPT generated plausible but incorrect information. This suggests that MyGPT may rely on general prior knowledge rather than accurately retrieving information from the linked sources. In contrast, uploading structured documents (PDFs) significantly improved response accuracy, probably because the model could directly extract relevant content.
These findings highlight the importance of explicitly providing MyGPT with domain-specific documentation when using it for software tutorials and documentation generation. Future work will explore whether structured summaries or extracted key sections from documentation further improve accuracy and reduce the need to upload full PDFs.

To further evaluate the capabilities of MyGPT, we tested its ability to generate a hyperparameter optimization tutorial using PHYSBO for neural networks configured with Keras~\cite{chollet2017deep}. 
We provided the following prompt:
\begin{quote}
    \textit{``Create a tutorial on optimizing hyperparameters in PHYSBO for neural networks configured using Keras.''}
\end{quote}
MyGPT generated a step-by-step tutorial covering key aspects of hyperparameter tuning with PHYSBO, including:

\begin{itemize}
    \item Installation of required packages.
    \item Definition of the hyperparameter search space, including learning rate, batch size, and number of hidden units.
    \item Implementation of a simple feedforward neural network in Keras, using the MNIST~\cite{726791} dataset as an example.
    \item Evaluation of hyperparameter configurations using PHYSBO.
    \item Execution of Bayesian optimization to determine the best hyperparameter set.
    \item Extraction and reporting of the optimal hyperparameter values.
\end{itemize}

The generated tutorial included relevant Python code snippets that demonstrate the integration of PHYSBO with Keras. The core of the tutorial focused on employing TS and Bayesian optimization strategies to maximize model performance.
The Python code generated by MyGPT for setting up and optimizing hyperparameters is uploaded to the ISSP data repository~\cite{PHYSBO-Gallary}.

To validate the accuracy and effectiveness of the tutorial, we executed the generated code. The optimization process successfully converged to an optimal configuration, achieving an accuracy of $0.9742$. The final hyperparameters were:

\begin{itemize}
    \item Learning Rate: 0.001
    \item Batch Size: 16
    \item Number of Hidden Units: 128
\end{itemize}

Additionally, intermediate optimization steps exhibited steady improvements, demonstrating the effectiveness of Bayesian optimization for hyperparameter tuning. The results confirm that MyGPT can generate functionally correct and executable tutorials when provided with structured reference materials such as official documentation and research papers.
However, continuous validation is still required to ensure accuracy, particularly when integrating AI-generated outputs into software documentation workflows.

This demonstration has shown that MyGPT can effectively generate executable tutorials for PHYSBO-based hyperparameter optimization when structured reference materials are provided. While initial responses contained inaccuracies, uploading domain-specific documentation significantly improved accuracy. The results suggest that AI-driven tutorial generation can enhance software usability and accessibility, though continuous validation remains necessary. Future work should explore integrating structured prompt engineering techniques to further refine the quality of generated outputs.

\section{Development of a GUI application from Python scripts with the help of generative AI}
\label{sec:GUI}

\begin{figure}
\centering
\includegraphics{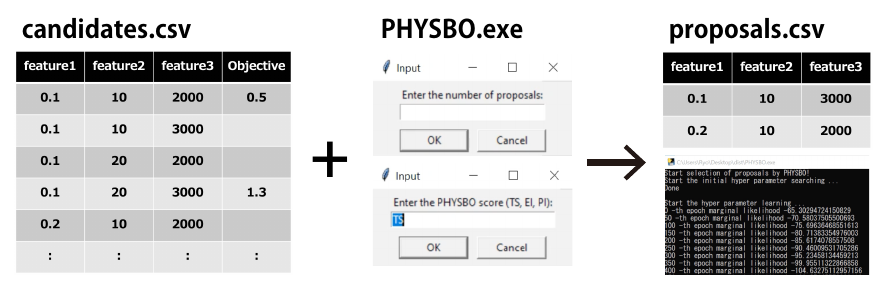}
\caption{Windows executable file for PHYSBO was developed with the help of ChatGPT. Based on the candidates.csv file, which includes experimental conditions and some objective function values, promising candidates expected to yield higher objective function values are proposed.}
\label{PHYSBO_EXE}
\end{figure}

Many Python scripts are available as open-source software (OSS) packages on GitHub, providing significant value for data analysis and machine learning tasks. However, if a proper Python environment is not set up, users cannot utilize these scripts on a local PC. To address this issue, converting Python scripts into Windows executable files is an effective way that allows them to be executed without any Python environment setup on Windows PCs.
This approach offers considerable advantages, such as enabling practical exercises in university and high school lectures, as well as corporate seminars, simply by sharing the executable files. Additionally, in product development, this method eliminates the need to set up Python environments on individual PCs. Another option for running analyses without setting up a Python environment is the use of cloud services. However, cloud-based solutions pose security risks when dealing with confidential data, as their use may not be feasible for such scenarios. By contrast, Windows executable files allow offline and local execution of analyses.  
One of their key benefits is their ability to ensure a high level of security against data breaches.

In this section, we introduce the development of a GUI application for the Bayesian optimization package PHYSBO (see Fig.~\ref{PHYSBO_EXE}), created with the help of ChatGPT. 
The original code used to convert to a Windows executable file was based on the PHYSBO code implemented in NIMO~\cite{Tamura2023,NIMO} (\verb|ai_tool_physbo.py|). 
In NIMO, the process begins by preparing \verb|candidates.csv|, which organizes all experimental condition candidates into a table. The rightmost column of \verb|candidates.csv| is designated to store the objective function values: if an experiment has already been conducted, the corresponding objective function value is entered, while for experiments not yet performed, the field remains blank. An example is shown in Fig.~\ref{PHYSBO_EXE}.
The experimental conditions with completed objective function values are used as training data, and the next experimental conditions to be tested are proposed from the blank fields. These proposed experimental conditions are output as \verb|proposals.csv|. It is important to note that the proposed explanatory variables are those predicted to maximize the objective function value. If the goal is to find experimental conditions that minimize the objective function value, this can be achieved by attaching a negative sign to the objective function values when storing them in \verb|candidates.csv|.

The Windows executable file was created through interactions with ChatGPT by following the steps below.

\vspace{10pt}
\noindent
(i) The prepared PHYSBO code was loaded into ChatGPT, and the following command was given: 
\begin{quote}
    \textit{``Convert the following script into an exe file so that it will work in a Windows environment."}.
\end{quote} 
In response, the Python code was generated, along with the following instructions. The PHYSBO code that was loaded can be found at \url{https://github.com/NIMS-DA/nimo/tree/main/nimo/ai_tools}.

\begin{quote}
To convert the given Python script into a Windows executable file, you can use PyInstaller. Below are the steps:
\begin{enumerate}

\item[1.]	\textbf{Install PyInstaller}: Ensure you have pyinstaller installed. You can install it via pip:

\verb|pip install pyinstaller|

\item[2.]	\textbf{Prepare the Script}: Make sure your script works correctly by itself when executed. It appears the provided script uses external dependencies such as numpy, physbo, etc. Ensure these are installed in your Python environment:

\verb|pip install numpy physbo|

\item[3.]	\textbf{Generate the Executable}: Run the following command in the terminal where the script is located:

\verb|pyinstaller --onefile PHYSBO.py|

\begin{enumerate}
\item[$\circ$]	\verb|--onefile| ensures the executable is packaged into a single file.
\item[$\circ$]	This will create a standalone .exe file in the dist folder.
\end{enumerate}

\item[4.]	\textbf{Locate the Executable}: After the command completes, the PHYSBO.exe file will be located in the dist directory.

\item[5.]	\textbf{Test the Executable}: Move the .exe to a folder with the required input files (e.g., \verb|candidates.csv|) and test it in a Windows environment.

\end{enumerate}

\end{quote}

\noindent
By following these instructions, the Windows executable file was successfully created, executed, and verified.

\vspace{10pt}
\noindent
(ii) We instructed ChatGPT as follows to modify the Python code so that the number of proposed candidates (\verb|num_proposals|) and the acquisition function (\verb|physbo_score|) could be specified as arguments:
\begin{quote}
\textit{``Please allow users to specify two arguments: }\verb|num_proposals| \textit{and} \verb|physbo_score|.".    
\end{quote}
 In response, the modified code was generated, and the following message was provided.

\begin{quote}

To modify the script to accept command-line arguments for \verb|num_proposals| and \verb|physbo_score|, we can use the argparse module. Here's the updated script:

\noindent
\textbf{Changes Made}:

\begin{enumerate}
\item[1.]	Added the argparse module to parse command-line arguments.
\item[2.]	Included \verb|--num_proposals| and \verb|--physbo_score| as required arguments.
\item[3.]	Updated the \verb|main()| function to read these arguments and pass them to the PHYSBO class.
\end{enumerate}

\noindent
\textbf{Usage}:

\noindent
After converting this script to an .exe file using pyinstaller, you can specify arguments like this:

\verb|PHYSBO.exe --num_proposals 5 --physbo_score TS|

\end{quote}

\noindent
This ensures flexibility for users to define the number of proposals and the scoring method.

\vspace{10pt}
\noindent
(iii) The Windows executable file created through the above steps requires \verb|num_proposals| and \verb|physbo_score| to be specified as arguments, and they need to be entered via the command prompt on a Windows computer. 
To enable GUI-based specification of these two arguments, we asked ChatGPT the following: 
\begin{quote}
\textit{``Please allow the user to specify two arguments,} \verb|num_proposals| \textit{and} \verb|physbo_score|,\textit{in the GUI."}    
\end{quote}
In response, the following message was provided.

\begin{quote}

\noindent
\textbf{Changes Made}:

\begin{enumerate}

\item[1.]	Replaced command-line arguments with a GUI using tkinter.
\item[2.]	Added input dialogs for \verb|num_proposals| and \verb|physbo_score| using simpledialog.
\item[3.]	Added error handling and user feedback using messagebox.

\end{enumerate}

\noindent
\textbf{How It Works}:
\begin{enumerate}
\item[1.]	When the script runs, users are prompted with GUI dialogs to input:

\begin{enumerate}
\item[$\circ$]	Number of Proposals.
\item[$\circ$]	PHYSBO Score (TS, EI, or PI).
\end{enumerate}

\item[2.]	If valid inputs are provided, the selection process proceeds, and a success message is displayed.
\item[3.]	Errors during input or execution show an error message.
This ensures the script is user-friendly in a Windows environment.
\end{enumerate}

\end{quote}

\noindent
Although we did not explicitly mention to ChatGPT that there are three types of PHYSBO scores, ChatGPT automatically suggested a feature to allow users to select among three acquisition functions.

\vspace{10pt}
\noindent
(iv) To make the application more user-friendly, we aimed to introduce a feature that allows the user to select an input file equivalent to \verb|candidates.csv|. Therefore, we instructed ChatGPT as follows: 
\begin{quote}
\textit{``Please allow the user to specify an input file equivalent to “candidates.csv” in the GUI.".}    
\end{quote}
In response, the following message was generated.

\begin{quote}

\noindent
\textbf{Updates}:

\begin{enumerate}
\item[1.]	Added a file selection dialog using \verb|filedialog.askopenfilename| to allow the user to specify the input file.
\item[2.]	Updated the \verb|main()| function to use the selected file as the \verb|input_file|.
\end{enumerate}

\noindent
\textbf{How It Works}:
\begin{enumerate}
\item[$\bullet$]	When the script runs:
\begin{enumerate}
\item[1.]	A file dialog prompts the user to select an input file (\verb|candidates.csv| equivalent).
\item[2.]	The user specifies the number of proposals and PHYSBO score as before.
\item[3.]	The selected input file is used in the optimization process.
\end{enumerate}
\end{enumerate}

\end{quote}

\noindent
This enhancement makes the script more flexible for different input files.

\vspace{10pt}
\noindent
(v) In the application created above, only a file selection screen appeared, and no message was displayed. Therefore, we instructed ChatGPT as follows:
\begin{quote}
    \textit{``Output a message to ask the user for the name of the input file in the GUI."}.
\end{quote}
In response, the following message was generated.

\begin{quote}

\noindent
\textbf{Updates}:
\begin{enumerate}
\item[1.]	Added a message box using \verb|messagebox.showinfo| to notify the user to select an input file before the file dialog appears.
\item[2.]	This makes the process more user-friendly by explicitly prompting the user.
\end{enumerate}

\noindent
\textbf{How It Works}:
\begin{enumerate}
\item[1.]	A message box pops up asking the user to select the input file.
\item[2.]	The file dialog then opens for the user to browse and select the required input file.
\end{enumerate}

\end{quote}

\noindent
This improvement enhances clarity and guides the user through the process.

\vspace{10pt}
\noindent
(vi) A feature to output predicted values for the test data through the GUI was implemented. In NIMO, setting the \verb|output_res| argument to True allows the output of predicted values, variance values, and acquisition function values for the test data. To enable this option in the GUI, we instructed ChatGPT as follows:
\begin{quote}
\textit{``Please allow users to specify whether} \verb|output_res| \textit{should be False or True in the GUI.". }
\end{quote}
In response, the following message was generated.

\begin{quote}
\noindent
\textbf{Updates}:
\begin{enumerate}
\item[1.]	Added a \verb|messagebox.askyesno| dialog to ask the user whether they want to enable \verb|output_res|.
\item[2.]	The user's choice (True for Yes, False for No) is passed to the PHYSBO class.
\end{enumerate}

\noindent
\textbf{How It Works}:
\begin{enumerate}
\item[1.]	After selecting the input file and entering other parameters, the user is prompted with:
Do you want to export prediction results?
\item[2.]	Clicking ``Yes'' sets \verb|output_res| to True.
\item[3.]	Clicking ``No'' sets \verb|output_res| to False.
\end{enumerate}
\end{quote}

\noindent
This allows users to choose whether to export the prediction results, directly via the GUI.
With the steps above, PHYSBO.exe has been completed and is publicly available on \url{https://isspns-gitlab.issp.u-tokyo.ac.jp/physbo-dev/physbo-gallery/-/tree/master/tools/physbo_exe}.

\vspace{10pt}
\noindent
(vii) Finally, to create a manual for the developed executable application, we instructed ChatGPT as follows: 
\begin{quote}
\textit{``Please create a manual for the GUI application created."}.    
\end{quote}

In response, the manual was generated along with the following message:
\begin{quote}
\noindent
\textit{``The user manual for the PHYSBO GUI Application has been created. It includes instructions for setup, operation, and troubleshooting. If you need modifications or additional details, feel free to ask!"}.
\end{quote}

A video demonstrating the execution of PHYSBO.exe can be found as Supplementary Movie 1. The manual created by ChatGPT could be used with almost no modifications. In addition, the manual for preparing the candidates.csv file was generated by loading the NIMO manual into ChatGPT. The two manuals were then merged and slightly modified by ChatGPT, and the result is available at \url{https://isspns-gitlab.issp.u-tokyo.ac.jp/physbo-dev/physbo-gallery/-/tree/master/tools/physbo_exe}. Since ChatGPT can easily perform language translation for the manuals, it becomes clear that the creation of both the Windows executable file and the manuals can largely be automated by generative AI. To prevent the Windows executable file from becoming too large, we used miniconda to prepare a minimal Python environment before creating the executable file. 
For simple GUI applications like this, generative AI can be effectively utilized to easily create them, and we expect it to be useful when creating educational content.

\section{Summary}
\label{sec:Summary}
In conclusion, this paper introduces three applications of generative AI: Testing software, creating an AI tutor for software, and developing a GUI. 
These tasks usually require significant effort, but generative AI greatly reduces the required workload, as demonstrated in this paper.
While generative AI does not always produce correct answers, human verification, such as testing, is necessary. 
However, our study confirms that generative AI often provides accurate results. Therefore, the use of generative AI in education and research in materials science is expected to become essential in the near future.

The rapid progress of generative AI will likely make the content of this paper outdated within a few years. However, we anticipate that this paper will serve as a valuable record of the early applications of generative AI in materials science.

\section*{Acknowledgement(s)}
A part of this study is supported by Data creation and utilization-type MaTerial R\&D project (DxMT).


\end{document}